\documentclass[reprint,superscriptaddress,aps,pra,floatfix,showpacs]{revtex4-1}
\usepackage{amsmath,amsfonts,amssymb,natbib,subfig,epsfig}
\usepackage{graphicx}
\usepackage{bm,dcolumn,hyperref}

\newcommand{\ket}[1]{\left\vert #1 \right\rangle}
\newcommand{\bra}[1]{\left\langle #1 \right\vert}

\newcommand{\ketbra}[2]{\left\vert #1 \right\rangle \left\langle #2 \right\vert}
\newcommand{\abs}[1]{\left\vert #1 \right\vert}

\begin{document}
\title{Quantum linear amplifier enhanced by photon subtraction and addition}
\author{Ho-Joon Kim}
\affiliation{Texas A\&M
University at Qatar, Education City, P.O. Box 23874, Doha, Qatar}
\author{Su-Yong Lee}
\affiliation{Texas A\&M
University at Qatar, Education City, P.O. Box 23874, Doha, Qatar}
\author{Se-Wan Ji}
\affiliation{School of Computational Sciences, Korea Institute for Advanced Study, Hoegiro 87, Dongdaemun, Seoul 130-722, Korea}
\author{Hyunchul Nha}
\affiliation{Texas A\&M
University at Qatar, Education City, P.O. Box 23874, Doha, Qatar}
\affiliation{School of Computational Sciences, Korea Institute for Advanced Study, Hoegiro 87, Dongdaemun, Seoul 130-722, Korea}
\date{\today}

\begin{abstract}
A deterministic quantum amplifier inevitably adds noise to an amplified signal due to the uncertainty principle in quantum physics.
We here investigate how a quantum-noise-limited amplifier can be improved by additionally employing the photon subtraction, the photon addition, and a coherent superposition of the two, thereby making a probabilistic, heralded, quantum amplifier. We show that these operations can enhance the performance in amplifying a coherent state in terms of intensity gain, fidelity, and phase uncertainty. In particular, the photon subtraction turns out to be optimal for the fidelity and the phase concentration among these elementary operations, while the photon addition also provides a significant reduction in the phase uncertainty with the largest gain effect.
\end{abstract}
\maketitle

\section{Introduction}
A perfect, deterministic, amplification of a quantum field is fundamentally impossible since a certain level of noise is inevitably introduced due to the uncertainty principle. 
This may put a practical limitation in quantum measurement and also provide a crucial basis for secure quantum communications. 
For example, the distinguishability of two quantum states cannot be improved by amplifying the input signals. In another respect, a malicious party intervening in quantum cryptography cannot obtain perfect clones of an unknown quantum state under communication \cite{Dieks}. The quantum-noise limit was identified for both the phase-insensitive and the phase-sensitive linear amplifiers by Caves \cite{caves1982}.  The noise in the phase-insensitive linear amplifier (PILA) generally degrades the nonclassicality of a quantum state being amplified unlike the case of the phase-sensitive one (e.g., squeezer). This decoherence effect has been studied for single-mode \cite{Hong} and two-mode amplified fields \cite{Agarwal, Nha}.  
A quantum-limited PILA can be experimentally realized by injecting a signal field to a nondegenerate parametric amplifier with the idler field in a vacuum state. 
It was also shown recently that the same noise limit can be achieved by employing a linear-optics scheme with a feed-forward based on homodyne detection \cite{Josse}.

In contrast to a deterministic scheme, the quantum noise can be significantly reduced if one adopts a probabilistic amplifying scheme. 
The so-called ``noiseless" amplifier, though probabilistic, can have practical applications for quantum communications particularly when the probabilistic event of success can be heralded by some means. 
Much attention has recently been drawn to the implementation of such noiseless amplifiers. 
Ralph and Lund proposed a scheme based on the quantum scissor used between splitting a given input into $N$ fields and recombining them \cite{ralph2009}, of which working principle was experimentally demonstrated in \cite{Xiang,ferreyrol2010}. This scheme, however, requires an arbitrarily large number of interferometric settings to achieve a high-fidelity performance, only with a very low success probability particularly for a large-amplitude input state \cite{Croke}. Zavatta {\it et al.} instead used a sequence of photon addition and photon subtraction, i.e., the operation ${\hat a}{\hat a}^\dag$ on a weak coherent state $|\alpha\rangle$ to realize a specific gain $g=2$ \cite{zavatta2010}. 
Another interesting scheme, i.e., adding thermal noise to an input state followed by photon subtraction, was proposed in \cite{marek2010}, which was also experimentally realized in \cite{usuga2010}. 
In the latter two schemes, the elementary photonic operations, subtraction ${\hat a}$ and addition ${\hat a}^\dag$, provide a key element for a probabilistic amplification. 
In fact, over the past years, the photon subtraction ${\hat a}$ proved to be a valuable resource for many applications in quantum informatics, e.g., improvement of teleportation fidelity \cite{opatrny2000}, distillation of entanglement \cite{browne2003,Takahashi}, manifestation of continuous variable nonlocality \cite{nha2004} and generation of the Schr\"odinger-cat-like states \cite{ourjoumtsev2006}. The photon addition ${\hat a}^\dag$ also provides similar advantages \cite{Yang} and can transform any classical state into a nonclassical state \cite{agarwal1991}.  Furthermore, the coherent superposition of those two operations was also proposed in the first order \cite{lee2010} and the high orders \cite{Kim1} of field operators for quantum-state engineering and fundamental tests. 

In this paper, we investigate how the quantum-limited PILA can be probabilistically enhanced by employing these elementary operations---photon subtraction ${\hat a}$, addition ${\hat a}^\dag$, and a coherent superposition $t{\hat a}+r{\hat a}^\dag$ \cite{lee2010}.  
We consider the amplification of coherent states and study the quality of the operations in terms of the gain, the fidelity, and the phase uncertainty of the output compared with the input. Compared to the scheme in \cite{marek2010,usuga2010}, the addition of thermal noise at the initial step is replaced by the use of quantum-limited PILA in our scheme while the photon subtraction at the last step can be replaced by the other photonic operations. 
The advantage of using the quantum-limited PILA over the thermal noise seems rather obvious because the former additionally gives a displacement effect, i.e., the increase of average amplitude, as well as the addition of noise \cite{Jeffers}.  
As the PILA and the photonic operations studied here are all within current experimental reach, our scheme represents another practical possibility for a noise-reduced quantum amplifier. 

This paper is organized as follows. In Sec. II, we briefly introduce the quantum-limited PILA and the probabilistic photonic operations. 
In Sec. III, we investigate and compare the effects of each probabilistic amplifier on input coherent states in terms of intensity gain, quantum fidelity, and the Holevo variance of phase. In Sec. IV, we summarize the main results.

\section{Quantum-limited phase-insensitive linear amplifier and probabilistic photonic operations}\label{Optical parametric amplifier and photon operations}
A PILA at the quantum-noise limit is characterized by the input-output relation as
\begin{eqnarray}
{\hat a}_G=\sqrt{G}{\hat a}_{\rm in}+\sqrt{G-1} {\hat v}^\dagger.
\label{eqn:map}
\end{eqnarray}
Here ${\hat a}_{\rm in}$ is an input mode, $\hat v$ is the vacuum mode characterizing the fundamental noise introduced via amplification, 
and ${\hat a}_{\rm G}$ is the amplified output at the intensity gain $G\geq 1$ \cite{caves1982}.  
This map can be practically implemented in a number of different ways. One is to inject a signal ${\hat a}_{\rm in}$ to a nondegenerate parametric amplifier with the idler mode in a vacuum state. Another is to split an input field ${\hat a}_{\rm in}$ via a beam splitter (BS) and then displace one output by the amount proportional to the outcome of homodyne measurement on the other output (feed-forward scheme with linear optics) \cite{Josse}.   

One can obtain the Glauber $P$ function of the amplified output corresponding to Eq. (\ref{eqn:map}) as
\begin{eqnarray}
P_G(\gamma)=\frac{1}{\pi(G-1)}\int\mkern-3mu d^2\beta P_{\rm in}(\beta)e^{-\frac{\left|\gamma-\sqrt G\beta\mkern2mu\right|^2}{G-1}},
\label{eqn:P-trans}
\end{eqnarray}
where $P_{\rm in}(\beta)$ is the $P$ function of the input state \cite{Nha}.
If the input mode is in a coherent state $\ket{\alpha}$ with $P_{\rm in}(\beta)=\delta^2(\beta-\alpha)$, the output state is given by 
\begin{eqnarray}
\rho_{G}=\frac{1}{\pi (G-1)}\int d^2\gamma\, e^{-\frac{\abs{\gamma-\sqrt{G}\alpha}^2}{G-1}} \ketbra{\gamma}{\gamma}
\label{eqn:PILA output}
\end{eqnarray}
in the coherent-state basis $|\gamma\rangle$. 
In Eq. (\ref{eqn:PILA output}), the $P$ function of the amplified state, $P_G(\gamma)=\frac{1}{\pi (G-1)}e^{-\frac{\abs{\gamma-\sqrt{G}\alpha}^2}{G-1}} $,  shows that the output distribution is broadened with the variance of $G-1$ (noise) while its center is displaced to $\sqrt{G}\alpha$ (amplification). 

Now we want to see how the output in Eq. (\ref{eqn:PILA output}) can be further amplified with reduced noise by applying some probabilistic operations. 
The conditional state after performing an operation ${\hat O}$ on $\rho_{G}$ is given by 
\begin{eqnarray}
\rho_{\rm op}=\frac{1}{N}{\hat O}\rho_G{\hat O}^\dagger
\label{eqn:conditional state}
\end{eqnarray}
where $N={\rm Tr} \left\{{\hat O}\rho_G{\hat O}^\dagger\right\}$ is a normalization constant. In this paper, we consider the operations 
${\hat O}={\hat a}$ (photon subtraction), ${\hat O}={\hat a}^\dag$ (photon addition), and ${\hat O}=t {\hat a}+r {\hat a}^\dag$ (coherent operation). 
For the case of ${\hat O}={\hat a}$, an initial state whose $P$ function is $P_{\rm in}(\gamma)$ is transformed to 
$\rho_{\rm op}\sim\int d^2\gamma P_{\rm in}(\gamma){\hat a}|\gamma\rangle\langle\gamma|{\hat a}^\dag=\int d^2\gamma |\gamma|^2P_{\rm in}(\gamma)|\gamma\rangle\langle\gamma|$, where the output $P$ function is the input distribution weighted by the intensity $|\gamma|^2$ that can make an amplification effect \cite{marek2010}. A similar argument can be given to the operation ${\hat O}={\hat a}^\dag$ for which the Glauber $P$ function is replaced by the Husimi $Q$ function to see the amplification effect. By studying the coherent operation ${\hat O}=t {\hat a}+r {\hat a}^\dag$, we can cover the whole range of photonic operations from subtraction ($r=0$) to addition ($r=1$).  (The condition $t^2+r^2=1$ is used without loss of generality.) 

In general, the $m$-photon subtraction operation, ${\hat a}^m$, can be approximately implemented by injecting an input to a beam splitter with high transmittance and then detecting $m$ photons at the auxiliary output mode \cite{usuga2010,ourjoumtsev2006,zavatta2009}. On the other hand, the $m$-photon addition operation, ${\hat a}^{\dag m}$, can be implemented by injecting an input to a nondegenerate parametric amplifier and detecting $m$ photons at the output idler mode \cite{zavatta2010,zavatta2009}. The coherent operation $t {\hat a}+r {\hat a}^\dag$ can be implemented using a Mach-Zehnder-type single-photon interferometer by erasing the which-path information on subtraction and addition \cite{lee2010}.
 
\section{Properties of the output states via probabilistic amplifiers}\label{Properties of the output states via probabilistic amplifiers}
In this section, we characterize the output state obtained by performing the operations  ${\hat a}^m$, ${\hat a}^{\dag m}$, and $t {\hat a}+r {\hat a}^\dag$, respectively, on the state in Eq. (\ref{eqn:PILA output}) amplified via a quantum-limited PILA. 
In particular, we investigate the intensity gain, the fidelity, and the phase uncertainty of the output state by each operation with respect to a target amplified state. 

The operations ${\hat a}$ and ${\hat a}^\dag$ commute with the phase-shift operation $e^{i\theta {\hat a}^\dag{\hat a}}$. That is, for ${\hat O}={\hat a}$ and ${\hat a}^\dag$,
\begin{eqnarray}
{\hat O}e^{i\theta {\hat a}^\dag{\hat a}}\rho e^{-i\theta {\hat a}^\dag{\hat a}}{\hat O}^\dag=e^{i\theta {\hat a}^\dag{\hat a}}{\hat O}\rho {\hat O}^\dag e^{-i\theta {\hat a}^\dag{\hat a}}.
\end{eqnarray}
It means that ${\hat a}$ and ${\hat a}^\dag$, respectively, yield the same effect on an input coherent state $\ket{|\alpha|e^{i\phi}}$ regardless of the phase $\phi$. This is not true with the coherent operation $t {\hat a}+r {\hat a}^\dag$, for which we therefore take into consideration the average over the whole range of input-state phase $\phi\in[0,2\pi]$ to evaluate performance in the following. We take $t$ and $r$ to be real ($t^2+r^2=1$) without loss of generality.
\subsection{Effective gain}
When the input field is in a coherent state $\ket{\alpha}$, the effective gain $G_e$ can be defined as
\begin{eqnarray}
\sqrt{G_e}\equiv\abs{\frac{{\rm Tr}\{{\hat a} \rho_{\rm op}\}}{\alpha}},
\end{eqnarray}
where ${\rm Tr}\{{\hat a} \rho_{\rm op}\}$ is the mean amplitude of the output state in Eq. (\ref{eqn:conditional state}).

\begin{figure}
\includegraphics[width=0.4\textwidth]{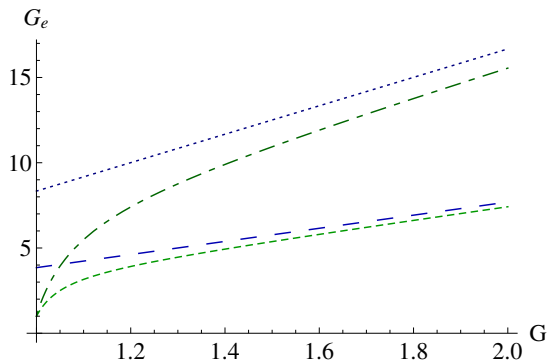}
\caption{(Color online) Effective gain $G_e$ versus the gain $G$ of the quantum-limited PILA followed by one-photon subtraction ${\hat a}$ (green dashed), two-photon subtraction ${\hat a}^2$(green dot-dashed), one-photon addition ${\hat a}^\dag$ (blue long-dashed), and two-photon addition ${\hat a}^{\dag2}$ (blue dotted), respectively, for the input coherent state with $|\alpha|=0.2.$}
\label{fig:effective gain}
\end{figure}

\begin{figure}
\includegraphics[width=0.4\textwidth]{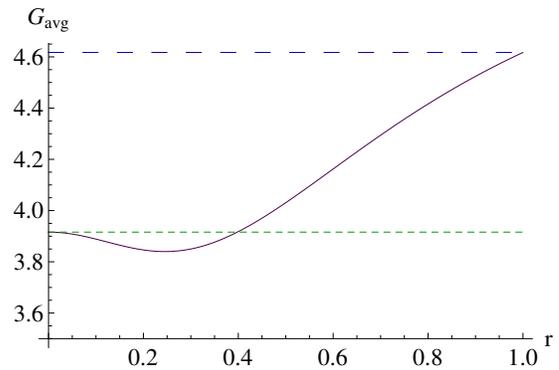}
\caption{(Color online) Effective gain $G_{\rm avg}$ averaged over the phase $\phi\in[0,2\pi]$ of input coherent states $\ket{|\alpha|e^{i\phi}}$ ($|\alpha|=0.2$) as a function of ratio $r$ in the coherent operation $t {\hat a}+r {\hat a}^\dagger$ following the quantum-limited PILA at gain $G=1.2$. The green dashed line (blue long-dashed line) shows the effective gain by one-photon subtraction ${\hat a}$ (one-photon addition ${\hat a}^\dag$).}
\label{fig:effective gain coh op}
\end{figure}

In Fig. \ref{fig:effective gain}, we show the effective gain $G_e$ achieved by first applying the quantum-limited PILA at gain $G$ and then further applying photon subtractions or photon additions. In general, $G_e$ turns out to be larger than $G$, which implies that both probabilistic operations successfully enhance the amplifying gain. As the number $m$ of operations ${\hat a}^m$ and ${\hat a}^{\dag m}$ increases, the effective gain becomes larger. In particular, the $m$-photon addition gives a higher effective gain than $m$-photon subtraction. Now, to see how the effect of coherent operation $t {\hat a}+r {\hat a}^\dagger$ fits into those of extremal operations ${\hat a}$ and ${\hat a}^{\dag}$, we plot the effective gain, averaged over all phases $\phi\in[0,2\pi]$ of input coherent states $\ket{|\alpha|e^{i\phi}}$, as a function of $r$ in Fig. \ref{fig:effective gain coh op}. 
We see that the coherent operation also enhances the effective gain overall for all values of ratio $r$, however, the effective gain does not increase monotonically with $r$ in general. The best operation turns out to be the photon addition ($r=1$) in view of the effective gain.   

\subsection{Output fidelity}
In the above, we have seen that all the considered photonic operations enhance the effective gain over the deterministic amplifier and that the photon addition among them gives the largest gain.
However, the intensity gain is not necessarily a good figure of merit for assessment of an amplifier. 
Thus we now consider the quantum fidelity between each output state and a target coherent state with amplitude $\sqrt{G_e}\alpha$, where $\alpha$ is an input amplitude. 
That is, we investigate the quantity 
\begin{eqnarray}
F\equiv \sqrt{\bra{\sqrt{G_e}\alpha}\rho_{\rm op}\ket{\sqrt{G_e}\alpha}}
\end{eqnarray}
for each operation and the results are shown in Fig. 3. 
\begin{figure}
\includegraphics[width=0.4\textwidth]{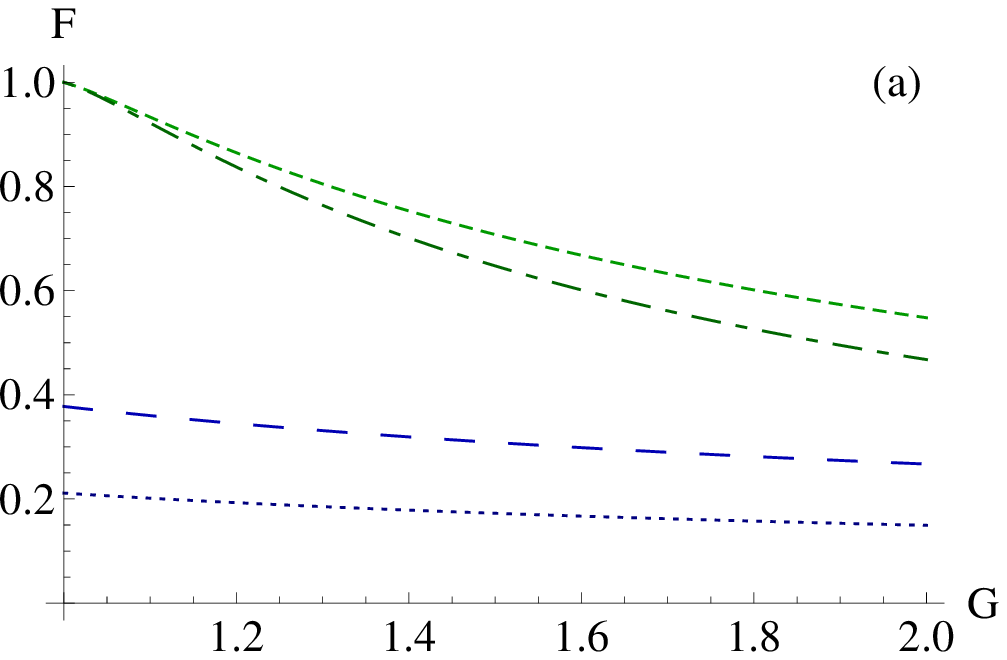}
\includegraphics[width=0.4\textwidth]{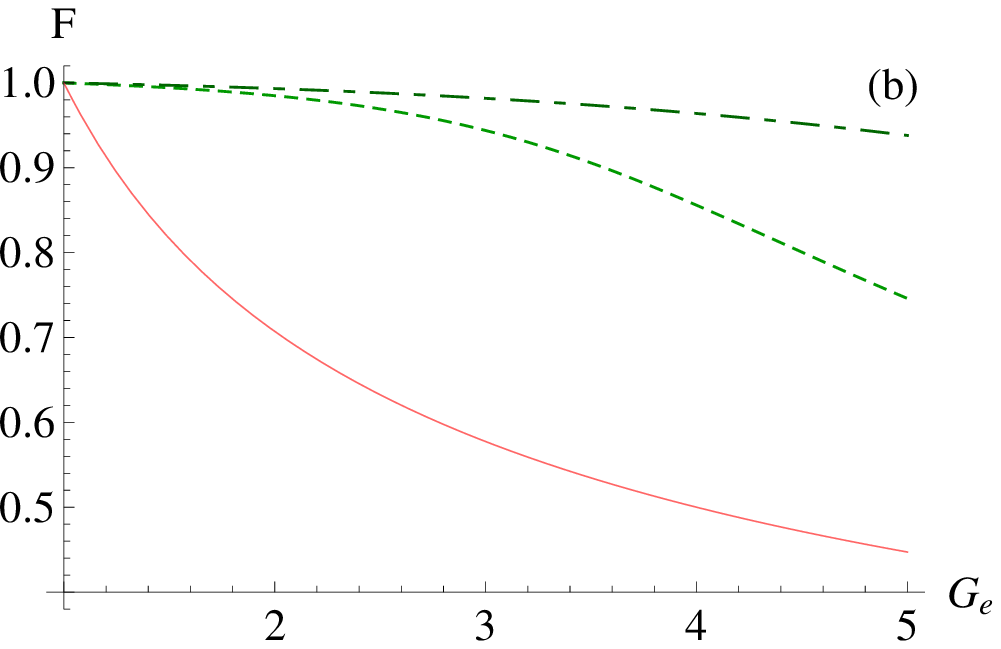}
\caption{(Color online) (a) Output fidelity $F$ versus the gain $G$ of the quantum-limited PILA followed by one-photon subtraction ${\hat a}$ (green dashed), two-photon subtraction ${\hat a}^2$(green dot-dashed), one-photon addition ${\hat a}^\dag$ (blue long-dashed), and two-photon addition ${\hat a}^{\dag2}$ (blue dotted), respectively, for the input coherent state with $|\alpha|=0.2$. (b) Output fidelity $F$ versus the effective gain $G_e$ for the subtraction schemes ${\hat a}$ (green dashed) and ${\hat a}^2$ (green dot-dashed), and the PILA alone (red solid).}
\end{figure}

We see that the output fidelity generally decreases with the amplifying gain for both operations. 
Furthermore, the fidelity appears to decrease with the number $m$ of operations ${\hat a}^m$ and ${\hat a}^{\dag m}$, for a given $G$, unlike the case of intensity gain in Fig. 1. 
However, if the fidelity is shown as a function of the effective gain $G_e$ in Fig. 3(b), we see that the output state has a better quality with more operations $m$ for ${\hat a}^m$.  For comparison, the fidelity for the case of using the PILA alone is also plotted, which is generally lower than that of the subtraction scheme. 
Within the same number of photon operations, photon subtraction operation gives a higher target fidelity than the photon addition scheme. This is because the photon addition operation always maps an initial classical state to a nonclassical state \cite{agarwal1991}, thereby significantly degrading the similarity between the target coherent state and the output state. 

In Fig. 4, we compare the scissor scheme \cite{ralph2009,Xiang} and the photon subtraction scheme in terms of the output fidelity and the success probability. 
The target effective gain is set to be $G_e=2$ for each scheme. 
For a small coherent input ($\alpha=0.2$), the scissor scheme has a slightly higher fidelity than the subtraction scheme using even a single scissor ($N$=1) with a considerable success probability.  
On the other hand, for a large coherent input ($\alpha=1$), the required number $N$ of scissors, in order to beat the output fidelity 
obtained by the subtraction scheme, is $N\ge3$, for which the success probability becomes lower than that of the subtraction scheme.  
For the case of subtraction scheme, which is implemented via a highly transparent beam splitter together with an on-off detector,
the success probability can be changed by adjusting the transmittance of the beam splitter, which of course affects the output fidelity.
For the plots of Fig. 4, we use the transmittance 0.99, which gives a modest level of success probability with a high output fidelity. 

\begin{figure}
\centering
\begin{tabular}{cc}
\epsfig{file=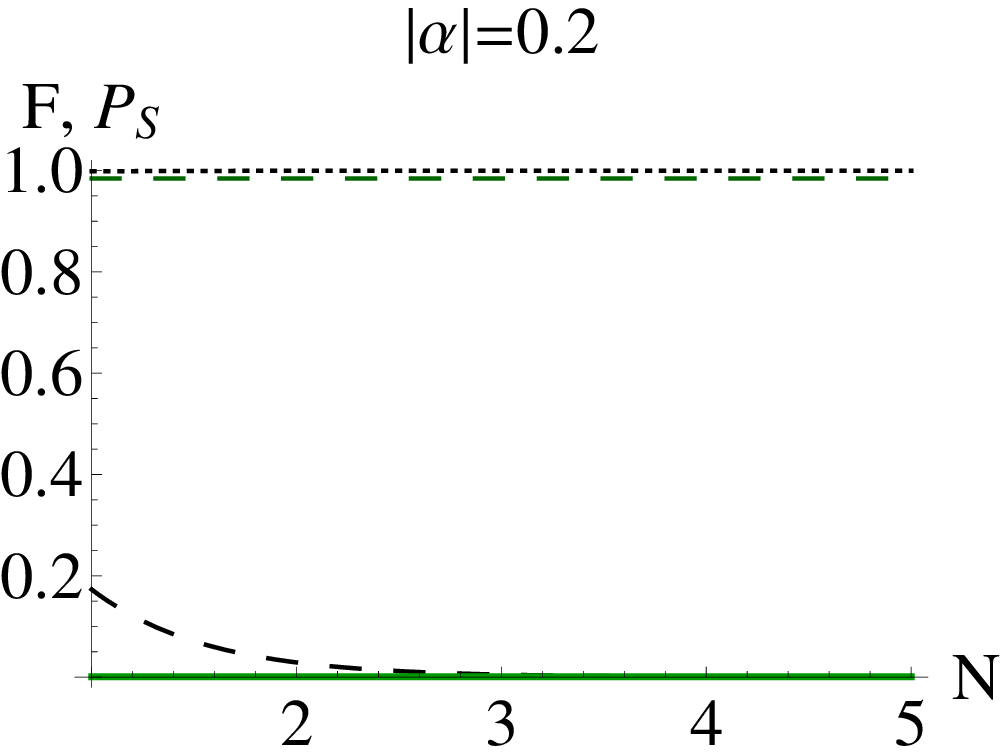,width=0.5\linewidth,clip=} & 
\epsfig{file=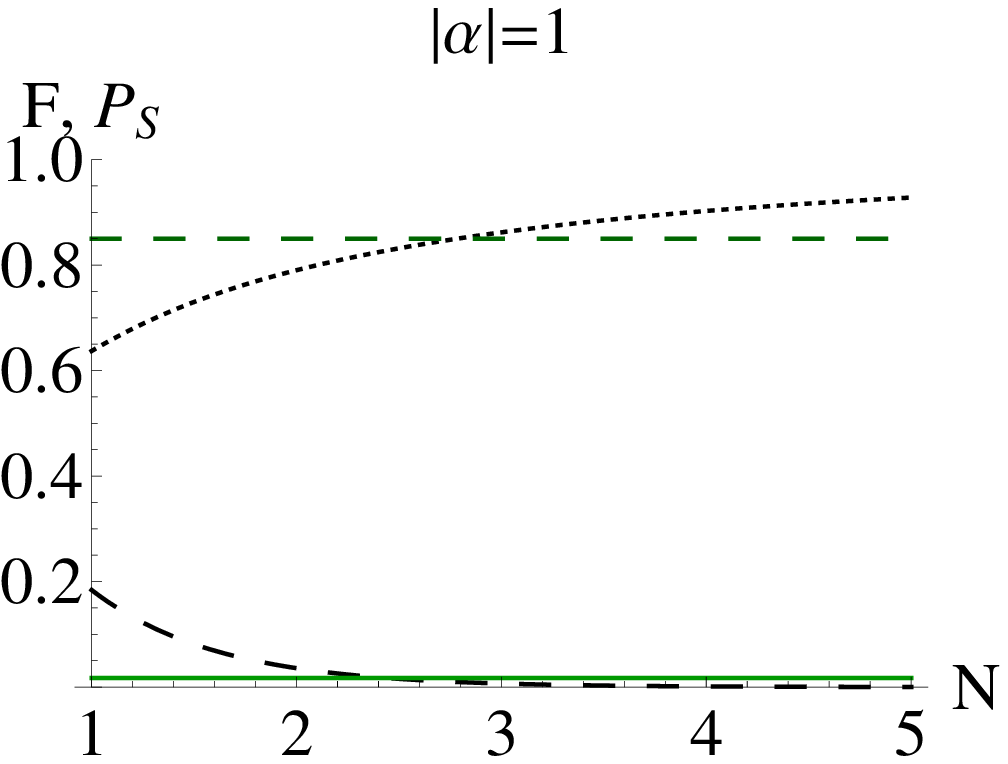,width=0.5\linewidth,clip=} \\
\epsfig{file=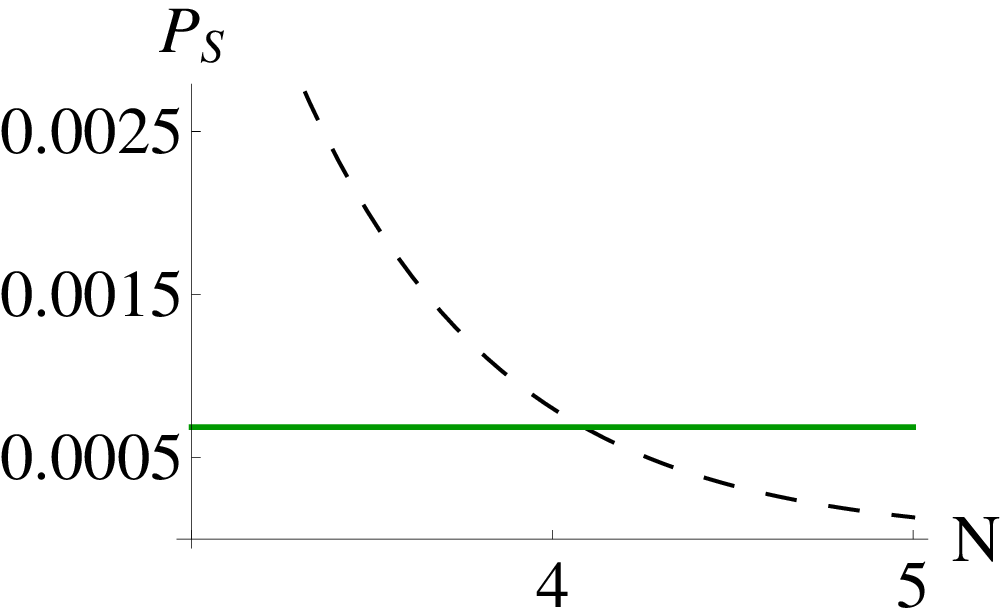,width=0.5\linewidth,clip=} &
\epsfig{file=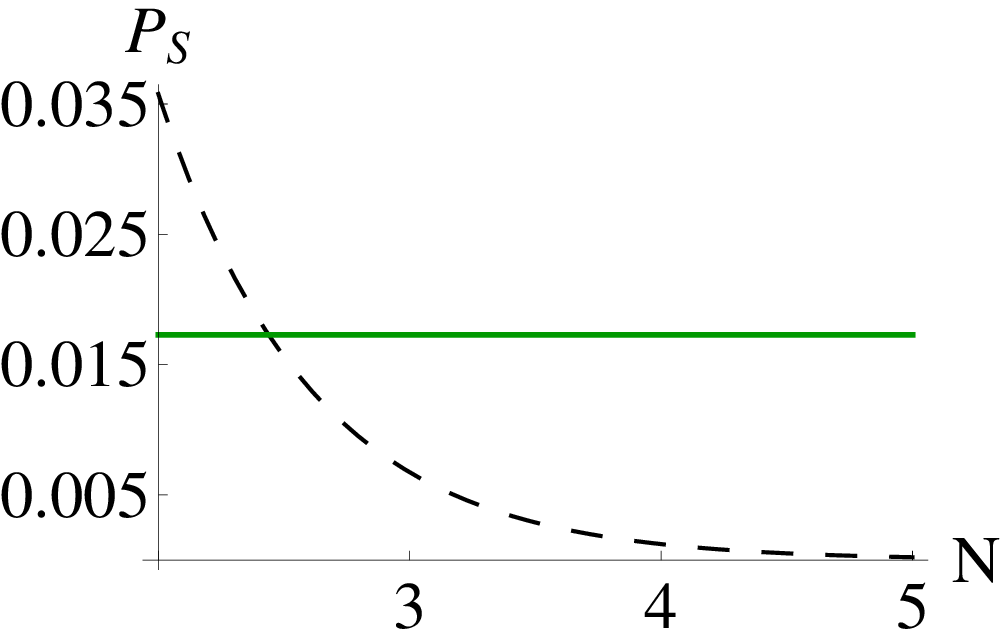,width=0.5\linewidth,clip=}
\end{tabular}
\caption{(Color online) Output fidelity $F$ (scissor scheme, black dotted; subtraction scheme, green dashed) and the success probability $P_s$ (scissor scheme, black dashed; subtraction scheme, green solid) to achieve the effective gain $G_e=2$. 
Left (right) panel: input amplitude $\alpha=0.2$ (1). Bottom figures represent the magnified view of the success probability $P_s$ shown above. $N$ is the number of the used scissors by splitting a given input into $N$ fields and recombining them \cite{ralph2009}. For the case of subtraction scheme, which may be implemented via a highly transparent beam splitter together with an on-off detector for heralding of success, the plots are made with the choice of the transmittance 0.99.}
\end{figure}

To better understand how each probabilistic operation modifies the phase-space profile of the coherent state, we plot the Wigner functions of the output states in Fig. 5.  
In general, the photonic operations push the Wigner distribution of the input field further away from the origin, hence enhancing amplification and phase concentration (see also Sec. III-C.) We see that the photon subtraction operation
does not significantly distort the initial Gaussian profile keeping the Wigner function positive definite at all points. On the other hand, the photon addition creates a negative region in the phase space with a more complex structure than the case of photon subtraction.  

\begin{figure}
\includegraphics[width=0.3\textwidth]{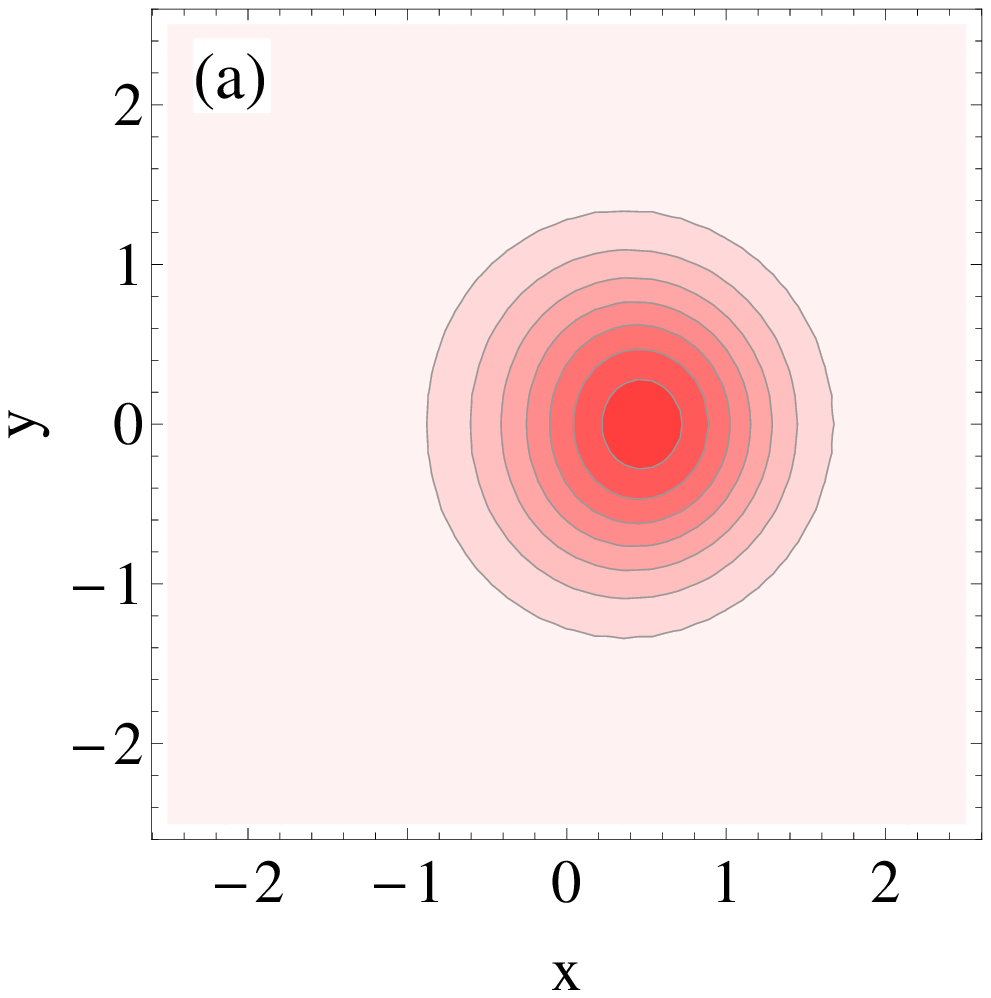}
\includegraphics[width=0.3\textwidth]{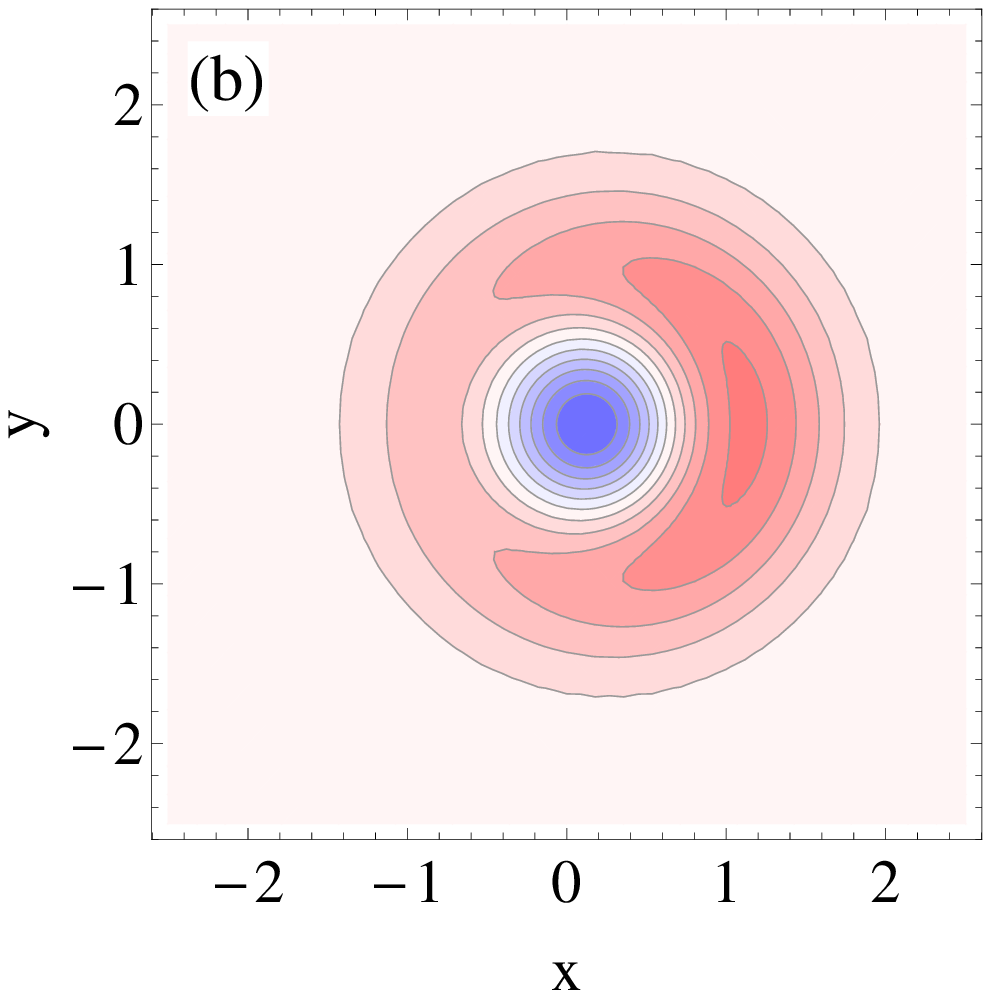}
\caption{(Color online) Wigner distribution of the output state by (a) one-photon subtraction ${\hat a}$ and (b) one-photon addition ${\hat a}^{\dag }$ following the quantum-limited PILA at gain $G=1.2$ for the input coherent state with $\alpha=0.2$. The red (blue) region designates the positive (negative) values of the Wigner function. The Wigner function takes all positive values in (a) whereas it shows a negative region around the origin in (b).}
\label{fig:Wigner distribution of the output field}
\end{figure}
\begin{figure}
\includegraphics[width=0.3\textwidth]{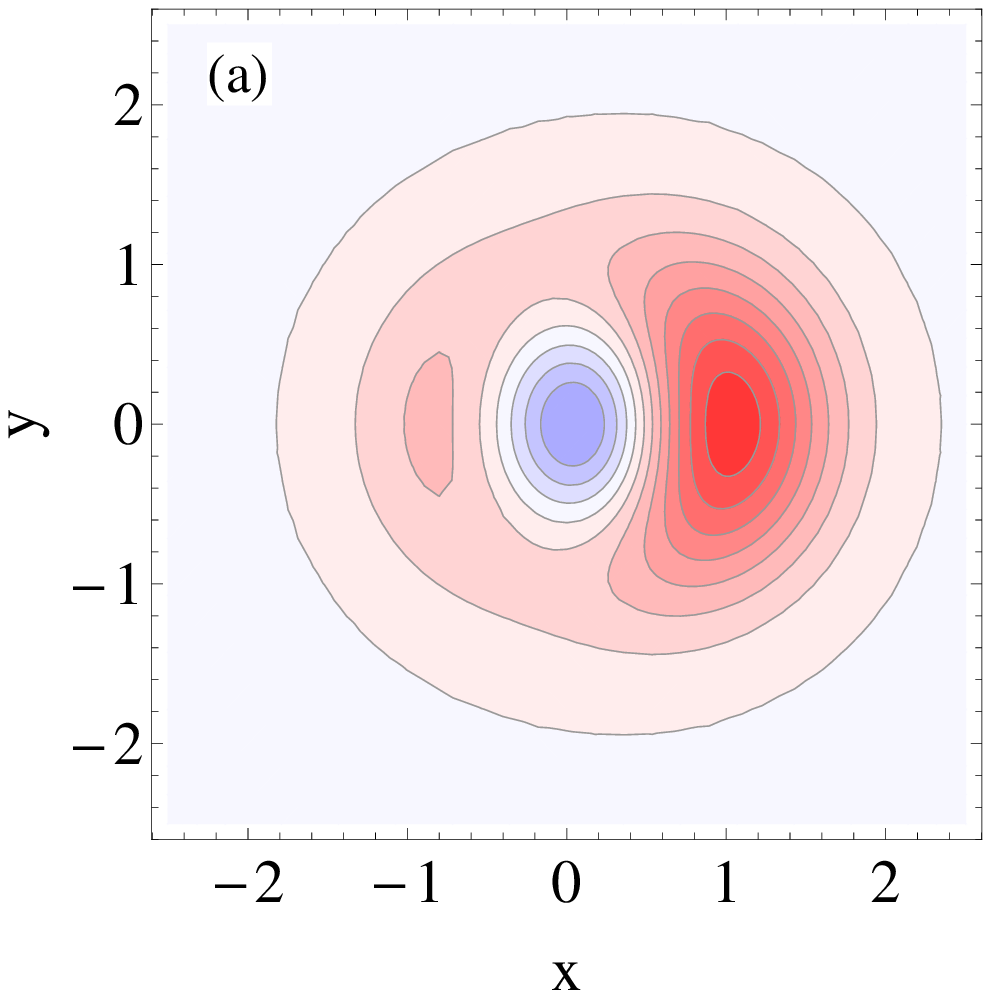}
\includegraphics[width=0.3\textwidth]{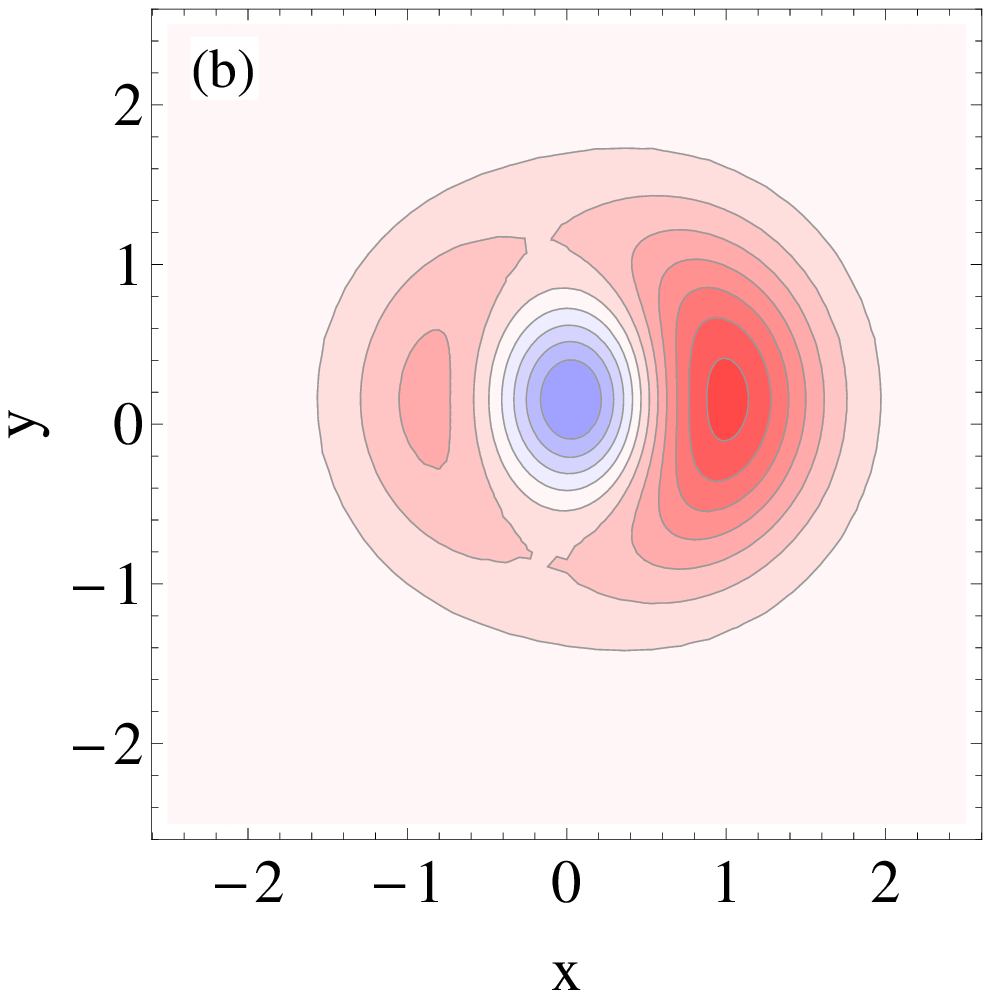}
\includegraphics[width=0.3\textwidth]{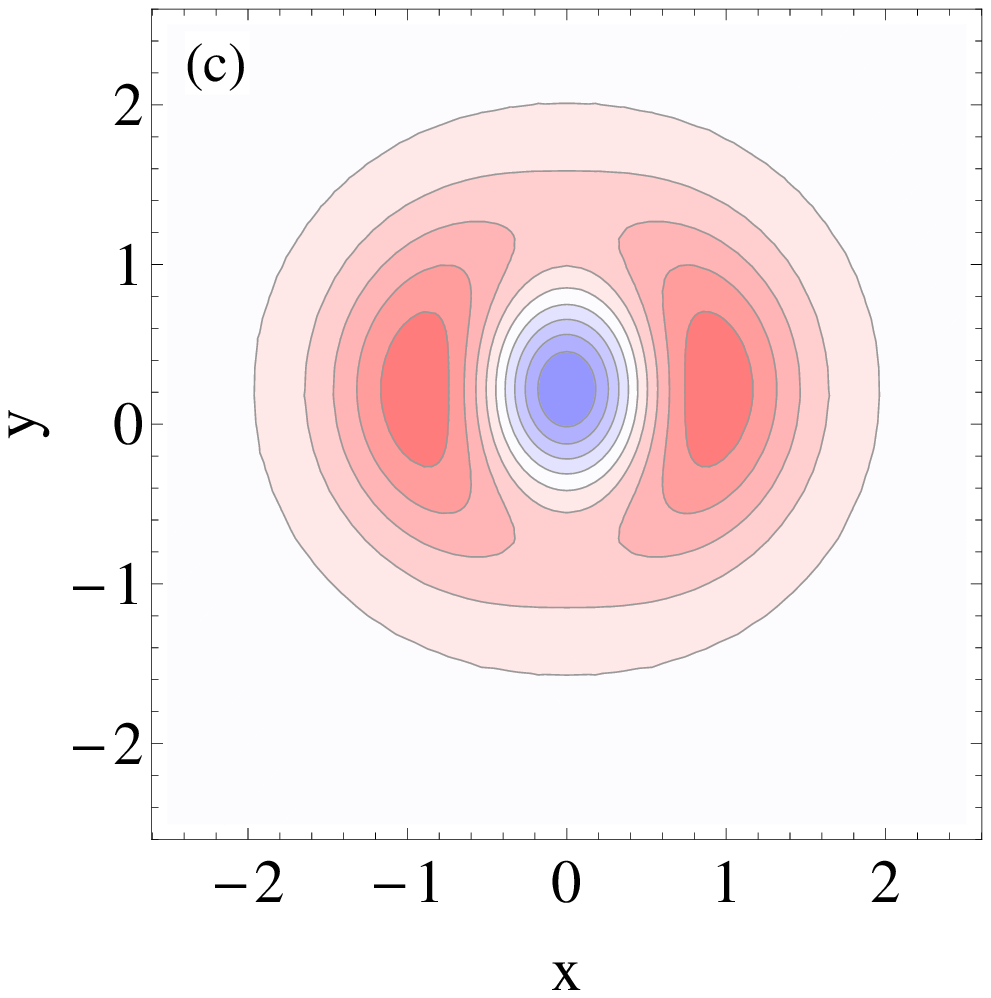}
\caption{(Color online) Wigner distribution of the output state by the coherent operation $t {\hat a}+r {\hat a}^\dagger$ ($r=\frac{1}{\sqrt{2}}$) following the quantum-limited PILA at gain $G=1.2$. The input coherent state $\ket{|\alpha|e^{i\phi}}$ with $\abs{\alpha}=0.2$ has the phase (a) $\phi=0$, (b) $\phi=\frac{\pi}{4}$, and (c) $\phi=\frac{\pi}{2}$, respectively. The red (blue) region designates the positive (negative) values of the Wigner function. In all plots, the Wigner function takes negative values in the central region around the origin.}
\label{fig: wigner dist coh op}
\end{figure}

In Fig. 6, we also plot the Wigner distribution of the output state by a coherent superposition operation $t {\hat a}+r {\hat a}^\dagger$ following the quantum-limited PILA. In this case, the output profile depends on the phase $\phi$ of the input coherent state $\ket{|\alpha|e^{i\phi}}$ as the operation $t {\hat a}+r {\hat a}^\dagger$ does not commute with the phase-shift operation. The coherent operation generally creates a negative region with a substantial non-Gaussianity of phase-space profile 
which can reduce the quantum fidelity between the output state and the target coherent state. To assess more precisely how the output fidelity changes as the ratio $r$ of the coherent operation $t {\hat a}+r {\hat a}^\dagger$ is changed , we plot the quantum fidelity averaged over the phase $\phi\in[0,2\pi]$ of input coherent states $\ket{|\alpha|e^{i\phi}}$ as a function of $r$ in Fig. 7. It is seen that the quantum fidelity shows a monotonic behavior with the ratio $r$ and that the photon subtraction ($r=0$) gives the best output fidelity among all operations.

\begin{figure}
\label{fig:effective fidelity}
\includegraphics[width=0.4\textwidth]{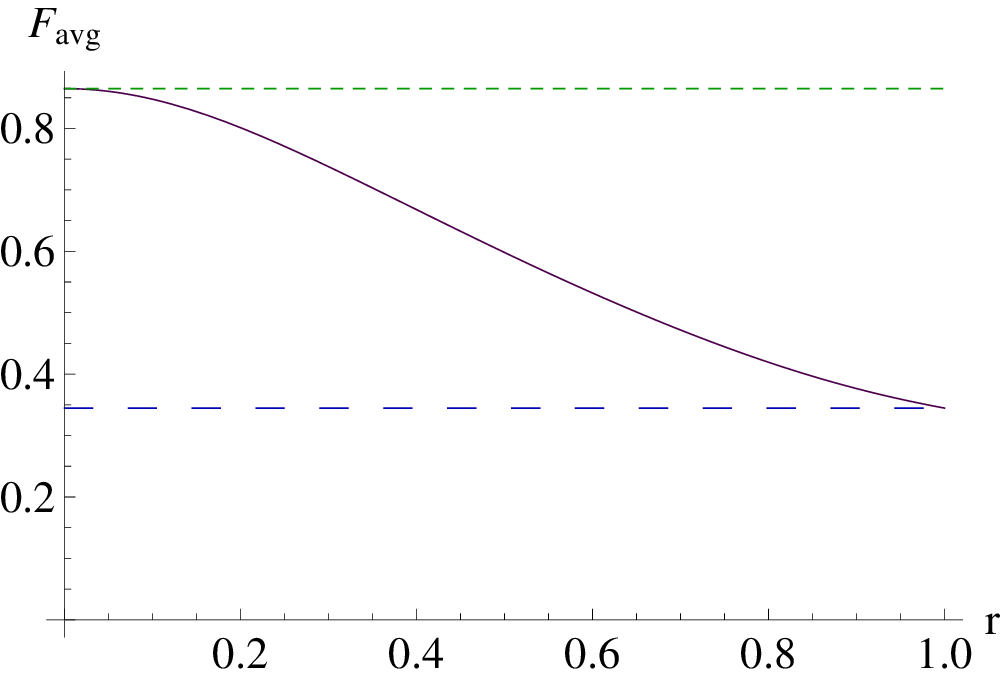}
\caption{(Color online) Output fidelity $F_{\rm avg}$ averaged over the phase $\phi\in[0,2\pi]$ of input coherent states $\ket{|\alpha|e^{i\phi}}$ as a function of $r$ by the coherent operation $t {\hat a}+r {\hat a}^\dagger$ following the quantum-limited PILA at gain $G=1.2$ ($\abs{\alpha}=0.2$). Green dashed line (blue long-dashed line) shows the output fidelity by one-photon subtraction ${\hat a}$ (one-photon addition ${\hat a}^\dag$).}
\label{fig:effective fidelity coh op}
\end{figure}

For the case of a very small coherent state, a rather high fidelity can be obtained even with a vacuum state. 
Thus, the quantum fidelity might not always be a right measure to look into. 
In the next section, we investigate the phase uncertainty of an output state to assess the quality of our schemes in another respect.

\subsection{Optical phase variance: Holevo variance}
In quantum communication using coherent states, the phase of the input state can play a role as an information content to deliver from one party to another. In such a case, the phase uncertainty of the output state after amplification becomes a subject of interest. Here we use the Holevo variance \cite{kholevo1982} to assess the quality of phase information, i.e., 
\begin{eqnarray}
V=\frac{1}{\abs{\mu}^2}-1,
\end{eqnarray}
where $\mu=\int_0^{2\pi} d\theta \, P(\theta) e^{i \theta}$ is the sharpness \cite{wiseman2009} corresponding to the probability distribution $P(\theta)$ of experimentally measured phase $\theta$. Using the ``canonical phase" \cite{pegg1989,leonhardt1995} the sharpness becomes $\mu_C=\sum_{n=0}^{\infty}\bra{n+1}\rho\ket{n}$, where $\ket{n}$ is the number state.

\begin{figure}
\includegraphics[width=0.4\textwidth]{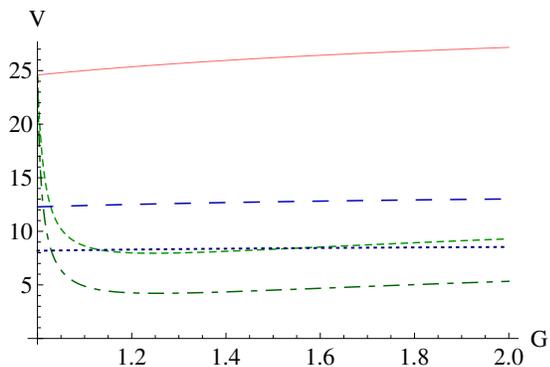}
\caption{(Color online) Holevo variance $V$ versus the gain $G$ of the quantum-limited PILA followed by one-photon subtraction ${\hat a}$ (green dashed), two-photon subtraction ${\hat a}^2$ (green dot-dashed), one-photon addition ${\hat a}^\dag$ (blue long-dashed), and two-photon addition ${\hat a}^{\dag2}$ (blue dotted), respectively, for the input coherent state with $|\alpha|=0.2$. Red (solid) line shows the variance from the quantum-limited PILA alone.}
\label{fig:holevo variance}
\end{figure}

In Fig. \ref{fig:holevo variance}, we plot the Holevo variance of the output state by photon subtractions and additions, respectively, following the quantum-limited PILA. Compared with the variance of the output state by the PILA alone (red solid line), the phase uncertainty is significantly reduced by each probabilistic operation, which is more pronounced with the number $m$ of the operations ${\hat a}^m$ and ${\hat a}^{\dag m}$. In particular, for the case of photon addition, the reduction of phase uncertainty is remarkable even without the PILA (case of $G=1$). This suggests that the photon addition alone can be useful to concentrate the phase information to some extent. 
To fully assess the performance of the photonic operations, we show in Fig. 9 the Holevo variance averaged over the phase $\phi\in[0,2\pi]$ of input coherent states $\ket{|\alpha|e^{i\phi}}$ as a function of $r$ by the coherent operation $t {\hat a}+r {\hat a}^\dagger$ following the quantum-limited PILA. We see that the coherent operation overall shows enhancement of phase concentration except for a certain narrow range of $r$. In general, the photon addition gives the best phase concentration at a low gain $G$ of the PILA, while the photon subtraction gives the optimal performance among the considered operations in a wide range of larger $G$.

\begin{figure}
\includegraphics[width=0.4\textwidth]{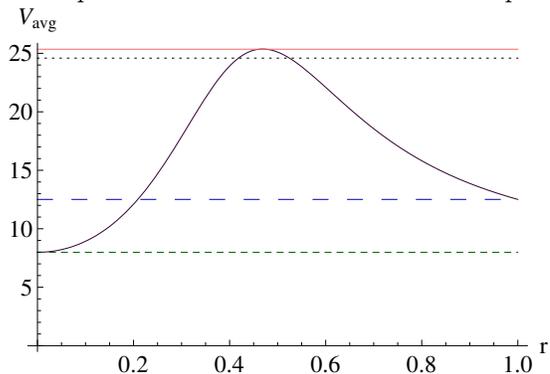}
\caption{(Color online) Holevo variance $V_{\rm avg}$ averaged over the phase $\phi\in[0,2\pi]$ of input coherent states $\ket{|\alpha|e^{i\phi}}$ as a function of $r$ by the coherent operation $t {\hat a}+r {\hat a}^\dagger$ following the quantum-limited PILA at gain $G=1.2$ ($\abs{\alpha}=0.2$). 
The green dashed line shows the case of one-photon subtraction ${\hat a}$, the blue long-dashed line the one-photon addition ${\hat a}^{\dag}$, the black dotted line the input coherent state, and the red straight line the PILA alone.}
\label{fig:holevo variance aver}
\end{figure}

\section{Conclusion}\label{conclusion}
In this paper, we have studied how a deterministic quantum amplifier can be enhanced by additionally applying some probabilistic operations. 
In particular, we have investigated the effects on an input coherent state of the quantum-limited PILA followed by the currently available photonic operations, i.e., photon subtraction, photon addition, and a coherent superposition of the two. It has been shown that these operations can enhance the performance of the deterministic amplifier in view of intensity gain, fidelity, and phase uncertainty. In particular, the photon subtraction turns out to be optimal among those operations in terms of fidelity and phase variance, while the photon addition also gives a significant phase concentration with the largest gain. As further applications, it will be interesting to study a nondeterministic, heralded, quantum cloner and an error correction scheme for continuous variables \cite{error} along a similar line to the present investigation. 

\emph{Note added.} After the completion of this work, we became aware of a related work \cite{Jeffers} where the photon subtraction following the quantum-noise-limited amplifier was considered.
\begin{acknowledgments}
This work is supported by NPRP Grant No. 08-043-1-011 from Qatar National Research Fund.
\end{acknowledgments}



\begin{thebibliography}{34}%
\bibitem{Dieks} W. K. Wootters and W. H. Zurek, Nature (London) \textbf{299}, 802 (1982); D. Dieks, Phys. Lett. A \textbf{92}, 271 (1982).
\bibitem{caves1982}C.~M. Caves,  Phys. Rev. D \textbf{26}, 1817 (1982).
\bibitem{Hong} B.~R.~Mollow and R.~J.~Glauber, Phys.~Rev.~{\bf 160}, 1076 (1967); C.~K.~Hong, S. Friberg, and L. Mandel, J. Opt. Soc. Am. B {\bf 2}, 494 (1985);
M.~Hillery and D. Yu, Phys. Rev. A {\bf 45}, 1860 (1992); G.~S.~Agarwal and K.~Tara, \emph{ibid.} {\bf 47}, 3160 (1993); H. Huang, S.-Y. Zhu, and M. S. Zubairy, \emph{ibid.} {\bf 53}, 1027 (1996).
\bibitem{Agarwal} G. S. Agarwal and S. Chaturvedi, Opt. Comm. {\bf 283}, 839 (2010).
\bibitem{Nha} H. Nha, G. J. Milburn, and H. J. Carmichael, New J. Phys. \textbf{12}, 103010 (2010).
\bibitem{Josse} V. Josse, M. Sabuncu, N. J. Cerf, G. Leuchs, and U. L. Andersen,  Phys. Rev. Lett. {\bf 96}, 163602 (2006). 
\bibitem{ralph2009}T.~C. Ralph and A.~P. Lund, \emph{Proceedings of the 9th International Conference on Quantum Communication, Measurement and Computing}, edited by A. Lvovsky (AIP, Melville, NY, 2009), pp. 155-160.
\bibitem{Xiang} G. Y. Xiang, T. C. Ralph, A. P. Lund, N. Walk and G. J. Pryde, Nature Photon. {\bf 4}, 316 (2010). 
\bibitem{ferreyrol2010}F. Ferreyrol, M. Barbieri, R. Blandino, S. Fossier, R. Tualle-Brouri, and P. Grangier,  Phys. Rev. Lett. \textbf{104}, 123603 (2010).
\bibitem{Croke} See also a scheme employing weak measurements by D. Menzies and S. Croke, arXiv: 0903.4181.
\bibitem{zavatta2010}A. Zavatta, J. Fiur\'a\v sek, and M. Bellini, Nature Photon. \textbf{5}, 52 (2011)
\bibitem{marek2010}P. Marek and R. Filip,  Phys. Rev. A \textbf{81}, 022302 (2010).
\bibitem{usuga2010}M.~A. Usuga, C.~R. M\"uller, P. Marek, R. Filip, C. Marquardt, G. Leuchs, and U.~L. Andersen, Nat. Phys. \textbf{6}, 767 (2010).
\bibitem{opatrny2000}
 T. Opatrn\'y, G. Kurizki, and D. G. Welsch,  Phys. Rev. A \textbf{61}, 032302 (2000).
\bibitem{browne2003} D. E. Browne, J. Eisert, S. Scheel,  and M. B. Plenio,  Phys. Rev. A \textbf{67}, 062320 (2003); S. L. Zhang and P. van Loock, {\it ibid.} \textbf{82}, 062316 (2010). 
\bibitem{Takahashi} H. Takahashi, J.S. Neergaard-Nielsen, M. Takeuchi, M. Takeoka,
K. Hayasaka, A. Furusawa, and M. Sasaki, Nature Photon. \textbf{4}, 178 (2010).
\bibitem{nha2004}H. Nha and H. J. Carmichael,  Phys. Rev. Lett. \textbf{93}, 020401 (2004); 
R. Garc\'a-Patr\'on, J. Fiur\'a\v sek, N.~J. Cerf, J. Wenger, R. Tualle-Brouri and P. Grangier, {\it ibid.} \textbf{93}, 130409 (2004).
\bibitem{ourjoumtsev2006}A. Ourjoumtsev, R. Tualle-Brouri, J. Laurat, and P. Grangier, Science \textbf{312}, 83 (2006); 
J.~S. Neergaard-Nielsen, B.~M. Nielsen, C. Hettich, K. M\o lmer, and E.~S. Polzik,  Phys. Rev. Lett. \textbf{97}, 083604 (2006); K. Wakui, H. Takahashi, A. Furusawa, and M. Sasaki, Opt. Exp. \textbf{15}, 3568 (2007)
\bibitem{Yang} Y. Yang and F.L. Li,  Phys. Rev. A \textbf{80}, 022315 (2009).
\bibitem{agarwal1991}G.~S. Agarwal and K. Tara,  Phys. Rev. A \textbf{43}, 492 (1991);  A. Zavatta, S. Viciani, and M. Bellini, Science \textbf{306}, 660 (2004).
\bibitem{lee2010}S. Y. Lee and H. Nha,  Phys. Rev. A \textbf{82}, 053812 (2010) ; S. Y. Lee, S. W. Ji, H. J. Kim,  and H. Nha, {\it ibid.} \textbf{84}, 012302 (2011).
\bibitem{Kim1} M. S. Kim, H. Jeong, A. Zavatta, V. Parigi, and M. Bellini,  Phys. Rev. Lett. \textbf{101}, 260401 (2008); 
A. Zavatta, V. Parigi, M. S. Kim, H. Jeong, and M. Bellini, {\it ibid.} \textbf{103}, 140406 (2009);  J. Fiur\'a\u sek,  Phys. Rev. A \textbf{80}, 053822 (2009).

\bibitem{zavatta2009}A. Zavatta, V. Parigi, M.~S. Kim, H Jeong, and M. Bellini,  Phys. Rev. Lett. \textbf{103}, 140406 (2009).

\bibitem{kholevo1982}A.~S. Kholevo, \emph{Probabilistic and
  Statistical Aspects of Quantum Theory} (North-Holland, Amsterdam, 1982)
\bibitem{wiseman2009}H.~M. Wiseman and G.~J. Milburn, \emph{Quantum Measurement and
  Control}, 1st ed. (Cambridge University Press, Cambridge, England, 2009).
\bibitem{pegg1989}D.~T. Pegg and S.~M. Barnett,  Phys. Rev. A \textbf{39}, 1665 (1989).
\bibitem{leonhardt1995}U. Leonhardt, J.~A. Vaccaro, B. B\"ohmer, and H. Paul,  Phys. Rev. A \textbf{51}, 84  (1995).
\bibitem{error} T. C. Ralph, Phys. Rev. A \textbf{84}, 022339 (2011).
\bibitem{Jeffers} J. Jeffers,  Phys. Rev. A \textbf{83}, 053818 (2011).
\end{thebibliography}

\end{document}